\RequirePackage{ifpdf}
\documentclass{JINST}
\usepackage{graphicx}
\usepackage{amsmath}
\usepackage{amssymb}

\def\eg{{\it e.g.}}
\def\fe{\ensuremath{^{55}\mathrm{Fe}}}
\def\pb{\ensuremath{^{210}\mathrm{Pb}}}
\def\po{\ensuremath{^{210}\mathrm{Po}}}
\def\rn{\ensuremath{^{222}\mathrm{Rn}}}

\title{A prototype low-background multiwire proportional chamber}

\author{Z.~Ahmed$^a$, M.~A.~Bowles$^b$, R.~Bunker$^b$, S.~R.~Golwala$^a$, D.~R.~Grant$^c$, M.~Kos$^b$, R.~H.~Nelson$^a$\thanks{Corresponding author.}, 
A.~Rider$^a$, R.~W.~Schnee$^b$, D.~Sotolongo$^a$, B.~Wang$^b$, and A.~Zahn$^a$\\
\llap{$^a$}Division of Physics, Math, and Astronomy\\
California Institute of Technology\\
 Pasadena, CA 91125, USA\\
\llap{$^b$}Department of Physics\\
Syracuse University\\
Syracuse, NY 13244, USA\\
\llap{$^c$}University of Alberta\\
  Edmonton, AB, T6G 2R3, Canada\\
  E-mail: \email{rhn@caltech.edu}}

\abstract{A prototype 
 multiwire proportional chamber (MWPC) was developed to demonstrate the feasibility of 
constructing a radiopure time projection chamber 
with MWPC track readout
to assay materials for 
alpha- and beta-emitting surface contaminants for future rare-event-search experiments as well as 
other scientific fields.  
The design features and assembly techniques described here are motivated by the position 
and energy resolution required to reconstruct alpha and beta tracks while efficiently 
rejecting backgrounds.
Results from a test setup using an $^{55}$Fe x-ray source indicate excellent operational stability and a near-ideal 
energy resolution of 15.8\% FWHM at 5.89\,keV and a gas gain of $\sim$10$^{4}$.}

\keywords{electron multipliers (gas), wire chambers, particle tracking detectors (gaseous detectors), time projection chambers (TPC)}

\begin{document}

\section{Introduction}
Future 
rare-event searches require unprecedented levels of radiopurity (see \eg~\cite{FormaggioBackgroundsAnnRev}).  
In particular, a troublesome source of contamination results from the noble-gas radioisotope \rn, which is sufficiently long lived to 
pervade laboratory spaces and can lead to the implantation of its long-lived \pb\ and \po\ daughters into or near sensitive detector surfaces.
Unfortunately, high-sensitivity detection of the $^{210}$Pb 46\,keV gamma ray from material surfaces is generally not feasible with HPGe detectors, in part due to its low branching fraction.
Furthermore, since these surface contaminants are chemically separated and hence out of equilibrium with the
photon-emitting isotopes in the parent $^{238}$U decay chain, direct detection of the $^{210}$Po alphas or betas from $^{210}$Pb or $^{210}$Bi
is necessary to establish the surface contamination level. Detection of surface betas is necessary for 
isotopes that can be detected only by the betas they emit ({\it e.g.}, $^3$H, $^{14}$C, $^{32}$Si, $^{63}$Ni, $^{90}$Sr, $^{106}$Ru, $^{113m}$Cd, $^{147}$Pm, 
$^{151}$Sm, $^{171}$Tm, $^{194}$Os, and $^{204}$Tl).
In addition to its importance for future rare-event searches, more sensitive detection of beta- and alpha-emitting surface 
contaminants would also benefit archeology, biology, climatology, environmental science,
geology, integrated-circuit quality control, and planetary science~\cite{schneeLRT2006}.

A screener intended for low-background detection of alpha and low-energy beta decays has several stringent design requirements:
 \begin{enumerate}
 \item A high-efficiency detection mechanism with good sensitivity to non-penetrating particles from sample surfaces ({\it e.g.}, a 50\,keV beta);
 \item A low energy threshold to maximize detection rates from isotopes that emit primarily low-energy betas ({\it e.g.}, $^{210}$Pb); 
 \item Sufficient energy resolution to permit identification of beta-emitting isotopes by their corresponding beta-spectrum endpoints and of isotopes that emit alphas, internal conversion electrons, or Auger electrons by their characteristic lines;
 \item Sufficient position information to allow efficient identification of the source location of contamination and rejection of detector or external backgrounds ({\it i.e.}, fiducialization);
 \item A well-shielded installation to reduce any external backgrounds that cannot be otherwise rejected; and
 \item Clean construction from radiopure materials to minimize intrinsic backgrounds.
\end{enumerate}
To achieve these goals, a gaseous time-projection chamber (TPC) is an ideal candidate.  Placing a sample directly in the detection 
medium allows for $\sim2\pi$ acceptance and very low energy thresholds as there is no dead layer (fulfilling requirements~1 and 2 above).  A multiwire proportional chamber~\cite{charpak} (MWPC) has sufficient position and energy 
resolution for imaging the ionized particle tracks when drifted across the chamber (allowing requirements~3 and 4 to be fulfilled).  
Passive shielding (requirement~5) of the TPC is reasonably straightforward,
and gases may be made sufficiently pure of intrinsic contaminants (partially fulfilling requirement~6).
These considerations led to the design of a TPC called the BetaCage~\cite{shutt_lrt2004,schneeLRT2006,ahmedLRT2010} with 
one MWPC near the sample to allow tagging of tracks starting from the sample, 
separated from a second MWPC by a distance large enough to range out betas up to 200\,keV in energy. 

The most challenging aspect of the low-background gaseous time-projection chamber is the construction of each radiopure MWPC. 
The MWPCs must be constructed from affordable materials with good radiopurity to reduce the amount of 
intrinsic backgrounds.  
They must be precision machinable to allow uniform wire gain, produce sufficiently low outgassing that electron collection is high and stable with time, 
and allow connections that produce sufficiently small noise that low thresholds may be achieved.
This article documents the design, construction, and performance of a prototype MWPC for use in the planned ultra-low-background screener, the BetaCage.  
Detailed simulations of expected backgrounds in the planned screener, incorporating the MWPC design described here, are summarized in~\cite{lrt2013ray} and will be fully described in a separate publication~\cite{BC-backgrounds}.


\section{MWPC design and construction}


Each MWPC consists of two cathode grids sandwiching a crossed anode grid.  
The prototype MWPC is a 3-cm thick by 3.8-cm wide frame enclosing a 40.1$\times$40.1\,cm$^2$ active area, 
while the MWPCs for the proposed screener will have 3-cm thick by 8.3-cm wide frames that each enclose an active area of 76.7$\times$76.7\,cm$^2$;
the screener's MWPC frames require additional width to help compensate for the increased strain imposed by their larger wire count.
The prototype uses P-10 to characterize its performance whereas the full screener will use neon-methane as neon has no long-lived radioisotopes (argon has $^{39}$Ar; a beta-emitter).  
To test the consistency of the design with the use of radiopure
construction materials, the prototype uses materials that are known to
be radiopure or available in radiopure form, with one exception: the
G10 printed circuit boards used for the readout wiring, for which
radiopure replacements will be discussed (\S\ref{wire:string}).
Construction of the planned screener will take place in a radon-abated
cleanroom~\cite{lrt2013richard} to ensure that radon daughters do not
contaminate critical surfaces.
This section details the choice of materials, the design requirements,
and the design and construction of the prototype MWPC.  

\subsection{Design requirements}\label{dsgn:req}

The ability to identify beta- or alpha-emitting isotopes imposes only relatively weak requirements on  
the detector energy resolution and hence the gain uniformity.
For alphas, a 10\% energy resolution is sufficient to prevent confusion between different alpha lines; other backgrounds in this energy region are expected to be negligible.
This 10\% energy resolution is more than sufficient to identify 
beta-decay spectra, since they are identified by their endpoints, which could never be measured better than 10\% with small numbers of counts.
The isotope 
\pb\ additionally has internal conversion lines in the 30-46\,keV range. 
At the (dominant) 30\,keV line, the intrinsic resolution  from carrier statistics is 5.7\%, so improvement of the contribution due to gain nonuniformity below 10\% is not especially helpful.
Gain uniformity must therefore be maintained to within $\sim$10\%.
The gain $G$ of a gaseous proportional chamber is given by the Diethorn formula~\cite{diethorn}
\begin{equation}
\ln G = \frac{\ln2}{\Delta V}\frac{\lambda}{2\pi\epsilon_0}\ln\frac{\lambda}{2\pi\epsilon_0aE_\mathrm{min}(\rho_0)\rho/\rho_0},
\label{die}
\end{equation}
where $\lambda$ is the linear charge density on the wire, $\epsilon_0$ is the vacuum permittivity, 
$a$ is the wire radius, $\rho_0$ is the density of the gas at STP, $\rho$ is the gas density, and $\Delta V$ 
and $E_\mathrm{min}(\rho_0)$ are empirical properties of the gas.  The parameter $\Delta V$ is the average potential required 
to produce at least one electron-ion pair, and $E_\mathrm{min}(\rho_0)$ is the minimum electric field required.  

\begin{figure}[!tb]
\begin{center}
\includegraphics[scale=0.45]{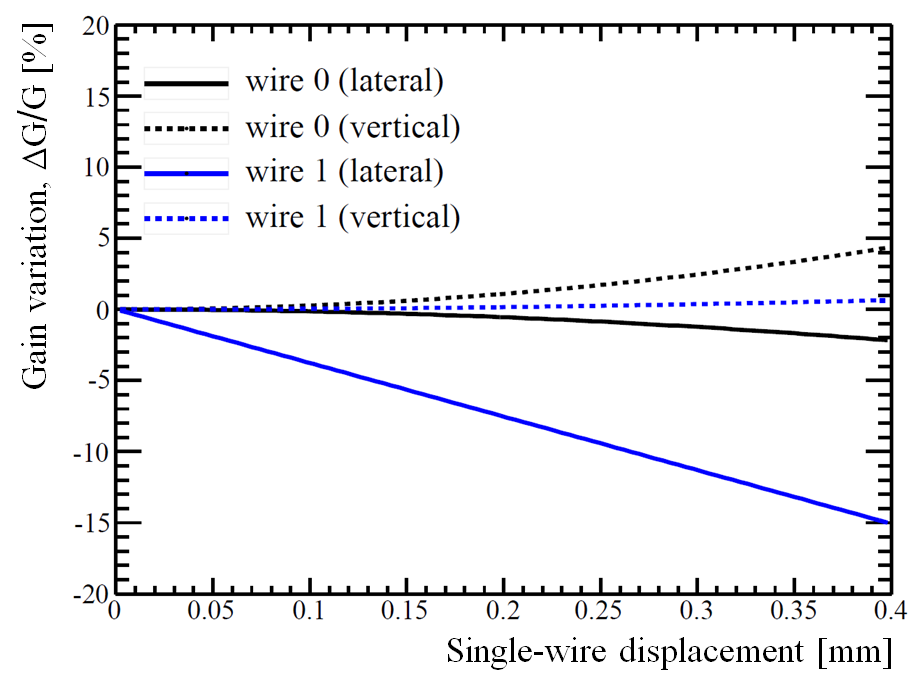}
\caption{Expected gain variations due to a single wire displacement, both on the displaced wire (wire 0, upper dotted and solid curves) and its nearest neighbor (wire 1, lower dotted and solid curves), where vertical and lateral displacements correspond to deviations out of and in the plane of the MWPC, respectively.}\label{disp}
\end{center}
\end{figure}

The expected variations in Eq.~\ref{die} come from voltage and pressure variations, wire displacements, 
and wire-diameter variations~\cite{pddc,sauli,erskine}.  The capacitance and potential of the anode with 
respect to the cathode grids sets $\lambda$.  A 5\% gain variation can result from a 2.5\,V (0.2\%) high-voltage 
source variation, a 1.5\% wire-diameter variation, or a 100\,$\mu$m (2\%) anode-wire displacement ({\it cf}.~Fig.~\ref{disp}). 
There is no avalanching at the cathode wires, so their positioning is not as important. Nevertheless,
as 
detailed in \S\ref{feed:thru}, the prototype's cathode and anode wires were positioned to within similar tolerances.   

\begin{figure}[!tb]
\begin{center}
\includegraphics[scale=0.8,clip]{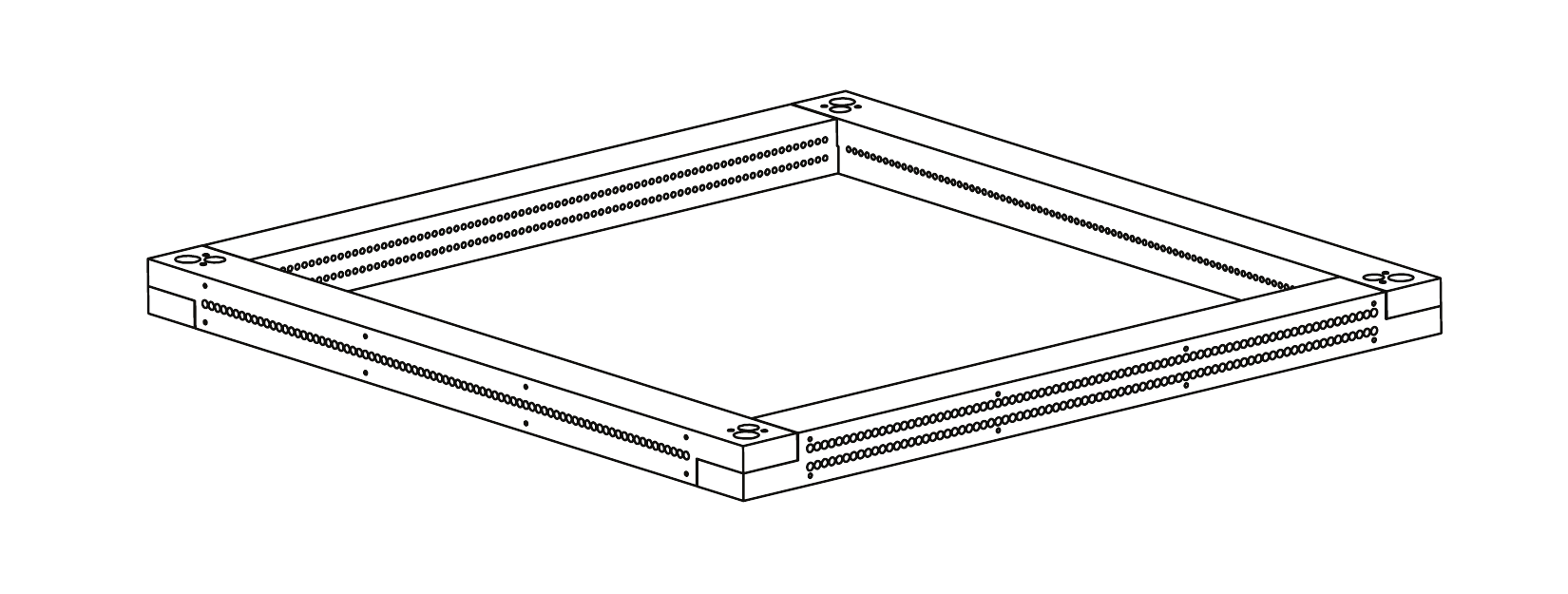}  
\caption{Mechanical drawing of the Noryl block-frame design for the prototype MWPC.  The 3-cm thick by 3.8-cm wide frame encloses a 40.1$\times$40.1\,cm$^2$ active area.  The anode and cathode layers are defined by the two sides with a single and double row of holes, respectively.  There are 79 wires per layer positioned by press-fit, spring-loaded feedthroughs to give a 5\,mm inter-wire and -layer pitch.}\label{fig:frame}
\end{center}
\end{figure}

\subsection{Block-frame design}
The MWPC frame is constructed from a high-density plastic called Noryl.\footnote{Originally developed by GE Plastics in 1966, Noryl is a proprietary amorphous blend of polyphenylene oxide (PPO) and polystyrene (PS)
that exhibits excellent hydrolytic and dimensional stability as well as good processability~\cite{sabic}. Screening using a variety of techniques indicates acceptable bulk contamination levels
for $^{238}$U, $^{232}$Th and $^{40}$K of $\leq$1, 3 and 5\,mBq/kg, respectively~\cite{lrt2013ray}.  While other plastics may be suitable, Noryl was attractive for these reasons as well as its low cost.  Acrylic, for example, is not as easily machined.}   
The frame was precision machined into four bars, two that hold the anode wires and two that contain the crossed cathode layers. 
Figure~\ref{fig:frame} shows a mechanical drawing of the prototype frame.  
Each layer contains 79 wires with a pitch of 5\,mm.  The anode-wire plane is also 5\,mm from each cathode-wire plane.
The four frame pieces were rough-cut, annealed, and finish-cut to a 20\,$\mu$m (or better) flatness, and then holes for the spring-loaded feedthroughs that hold the wires were drilled. The feedthroughs have a tapered shoulder in the middle for seating. 
The four pieces were arranged as shown in Fig.~\ref{fig:frame} with stainless-steel alignment pins and bolted together.  A 0.5~inch 
through-hole was drilled into each corner to facilitate connection to either a stand or 
fieldcage.  For the prototype frame, these holes were additionally used to bolt on copper-clad G10 sheets, the purpose of which is discussed in \S\ref{readout:elec}. 

\subsection{Spring-loaded feedthroughs}\label{feed:thru}
\begin{figure}[!tb]
\begin{center}
\begin{tabular}{cc}
\includegraphics[width=3.1in]{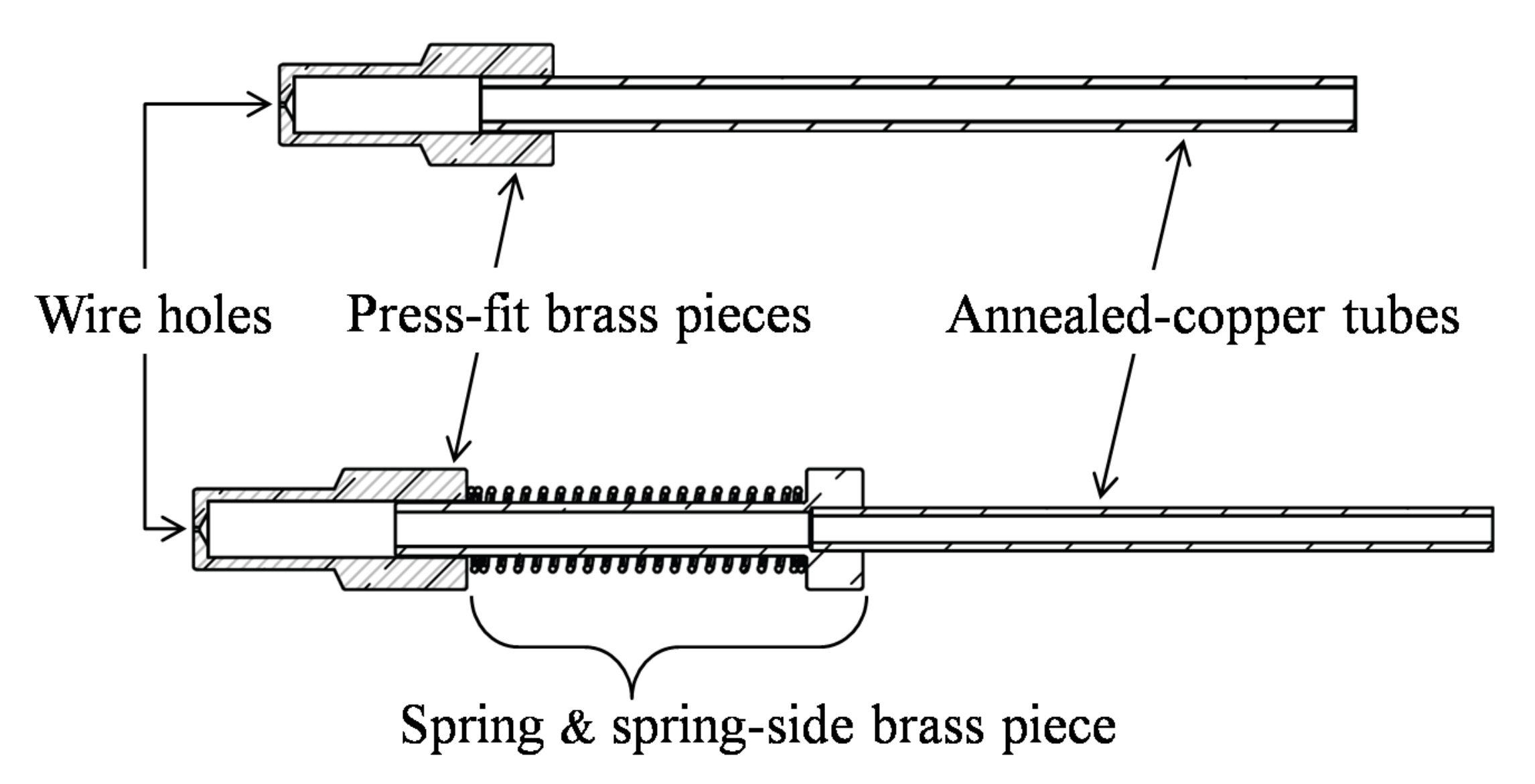}&
\includegraphics[width=2.52in]{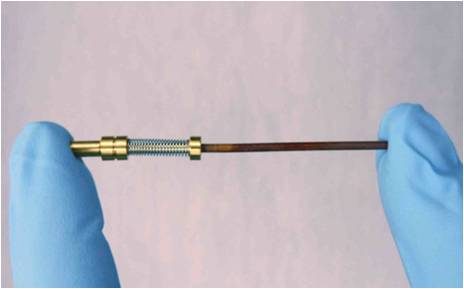}\\
a)&b)
\end{tabular}
\caption{ a) Mechanical drawings of the two feedthroughs used to position and tension each wire.  Stringing starts by threading the spring-side feedthrough (lower drawing) with a stainless-steel wire and crimping the copper tube (to hold the wire in place).  The feedthrough is then press fit into one side of the MWPC frame, and the wire is strung across the frame to the opposing side and threaded through a simpler, spring-less feedthrough (upper drawing), where it is tensioned and finally secured by crimping the second copper tube.  
b) Photograph of a spring-side feedthrough.}\label{fig.feed}
\end{center}
\end{figure}

To prevent gravitational sag, the wires are maintained at uniform tension through the use of custom-design feedthroughs. 
Each wire is held in place at one end by a spring-loaded feedthrough consisting of two precision-machined
brass pieces, a spring, and an annealed copper tube ({\it cf.}\ lower drawing in Fig.~\ref{fig.feed}a).  At the other end, the wire is held by a simpler, spring-less feedthrough consisting of a single brass piece and an annealed copper tube ({\it cf.}\ upper drawing in Fig.~\ref{fig.feed}a).  The wires are first threaded through the spring-side feedthrough and anchored by crimping its copper tube.  The anode (cathode) wires are then
threaded through the opposing, spring-less feedthrough, tensioned at 20\,g (200\,g) using a weight and pulley,
and finally secured by crimping the second copper tube.  In addition to gripping the wires,
the crimps establish electrical connections between the wires and copper tubes, which are used to instrument signal readout and provide high voltage.

As discussed in \S\ref{dsgn:req}, to prevent significant gain variations due to misplacement, a 100\,$\mu$m tolerance was specified for the positioning of each wire.
To achieve this, the feedthroughs were designed to be press-fit into the MWPC frame with a precision in position better than 25.4\,$\mu$m.
Further, each press-fit brass piece was designed with an anode (cathode) wire hole with a diameter of 101.6\,$\mu$m (177.8\,$\mu$m) centered to better than
25.4\,$\mu$m, thereby accurately guiding each 25.4-$\mu$m (127-$\mu$m) diameter stainless-steel anode (cathode) wire.\footnote{Gauge 50 (36) type 304 stainless-steel wire 
from the California Fine Wire Co. was used for the anode (cathode)~\cite{cafinewire}.} 
Following fabrication, the prototype MWPC frame and press-fit brass pieces were surveyed, indicating realized tolerances of 20 and 10\,$\mu$m, respectively.  
The resulting displacement is thus expected to be $<$45\,$\mu$m, well within the design specification described in \S\ref{dsgn:req} to maintain uniform gain.


All of the above elements (stainless-steel wire, brass pieces, copper
tubes, springs) can be readily made from radiopure stock.  Further, the wire can
be electropolished to remove surface radiocontaminants such as
$^{210}$Pb and $^{210}$Po~\cite{EPlrt2013,zuzel}.

\subsection{Wire stringing}\label{wire:string}
\begin{figure}[!tb]
\begin{center}
\begin{tabular}{cc}
\includegraphics[width=2.8125in]{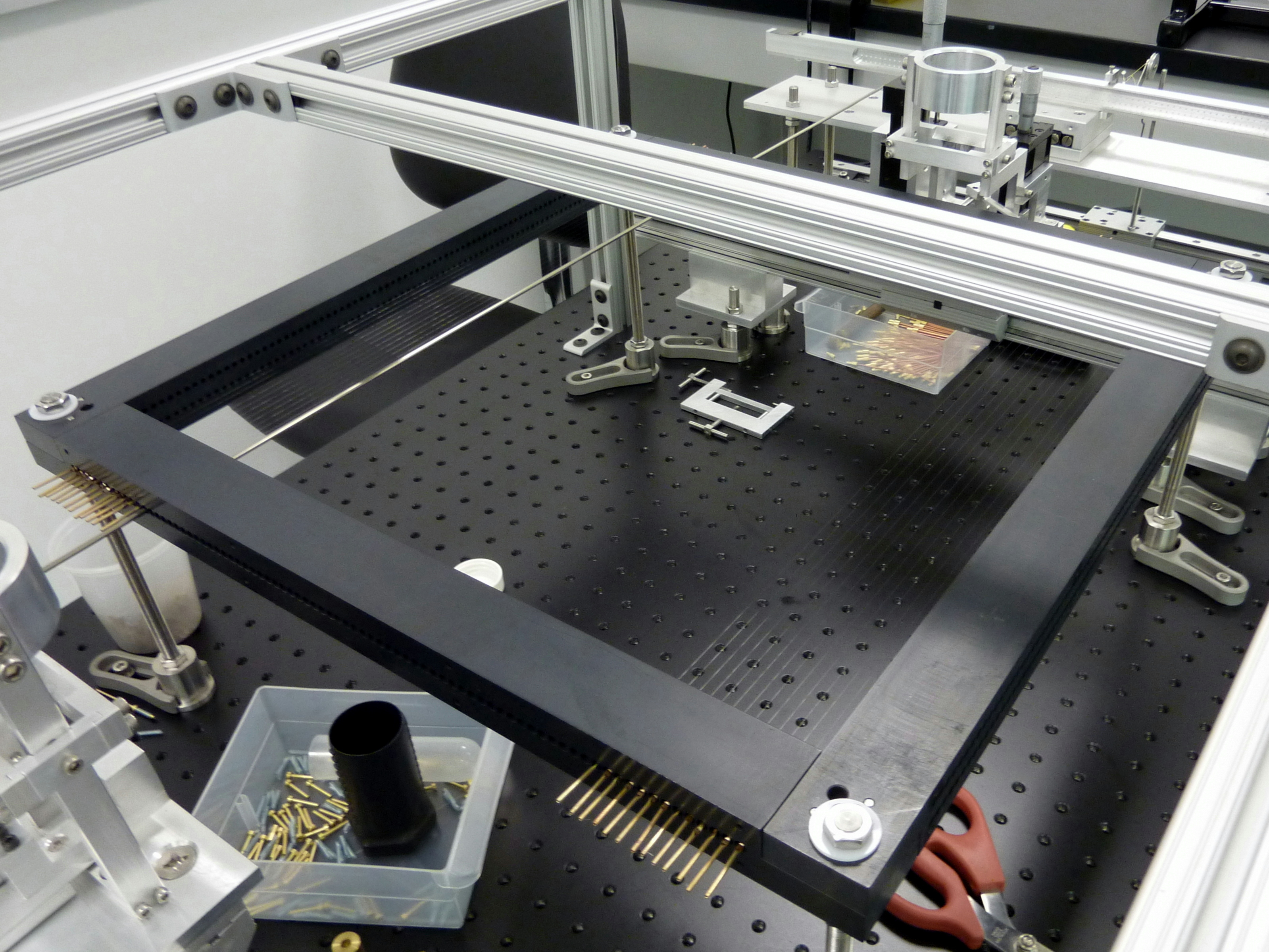}&
\includegraphics[width=2.8125in]{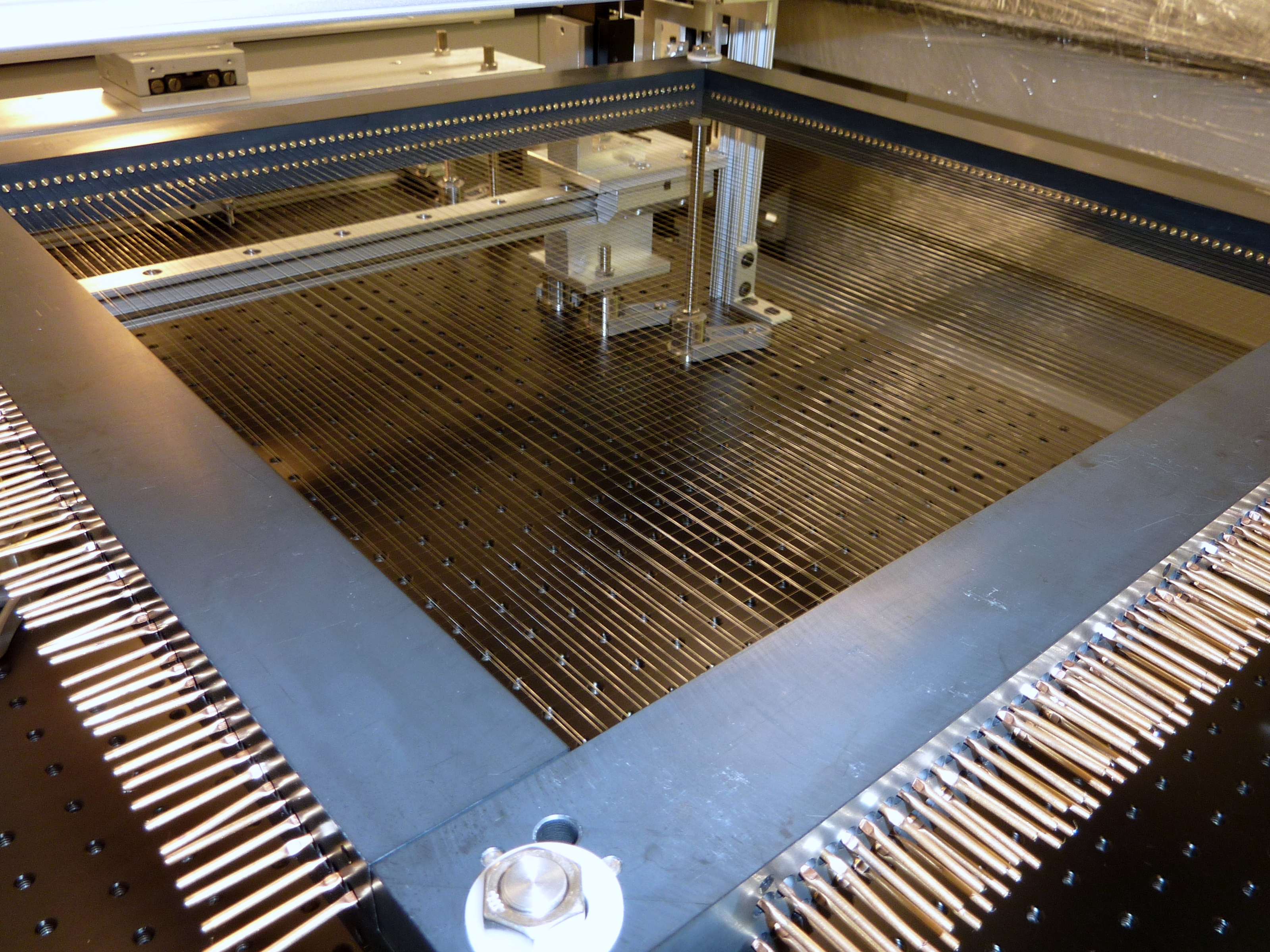}\\
a)&b)
\end{tabular}
\caption{a) Photograph of the setup used to string the prototype MWPC frame.  During the stringing process, the frame was secured to an optical table via threaded rods at each of the four corners and surrounded by a T-slotted aluminum frame to help protect the wires.  Custom-design fixtures for positioning a crimping tool were located on rails just outside opposing sides of the anode frame pieces.  An aluminum rod was used to unspool wire from the far side and through two opposing frame holes.  After anchoring the wire to the frame via the spring-side feedthrough (near side), it was tensioned using an aluminum bar attached to a low-friction slide (top of photo behind crimping fixture) and secured via the spring-less feedthrough.  The frame was rotated 90$^{\circ}$ in order to string the cathode layers.
b) Photograph of the fully-strung MWPC prototype. Assembly was completed by crimping fly-wires to the copper tubes along the outer edges of two adjacent sides of the frame.
The fly-wires were connected to custom circuit boards mounted directly to the frame.}\label{fig.frame.pic}
\end{center}
\end{figure}

The MWPC wires were strung by hand using a multi-step procedure (see also appendix~C in~\cite{ahmedthesis}).
As demonstrated in Fig.~\ref{fig.frame.pic}a, each wire was guided through opposing feedthrough holes using 
a long aluminum rod inserted through the holes to pull the wire back through the frame. 
The end of the wire at the spring-side was then threaded through the fully assembled 
spring-side feedthrough with the help of suction on the end of its copper tube (via a weak vacuum).  
With the threaded feedthrough held in place by a custom jig containing a pneumatic crimp tool,  
the feedthrough's copper tube was crimped, and the whole feedthrough was carefully press fit into the frame.
The opposite end of the wire was then inserted into the spring-less feedthrough, also with the help of suction.  Before tensioning the 
wire, the feedthrough was carefully press fit into the frame and a similar crimping jig was moved into place.  A custom-machined 
aluminum bar with the same hole pattern as the frame was affixed to a nearly frictionless slide.  A weight, hanging 
from the bar using a low-friction pulley, provided tension to the wire, at which point the spring-less feedthrough's
copper tube was crimped.  It took 6 minutes to string a single wire. The fully strung prototype MWPC is shown in Fig.~\ref{fig.frame.pic}b.

The final step in the MWPC assembly was to affix custom printed circuit boards (PCBs) to two edges of the frame.
Fly-wires crimped to the ends of the copper tubes were then attached to the PCBs, thereby establishing electrical connections from the MWPC wires to 
coaxial cabling (soldered directly to pads on the PCBs).  The coaxial cabling used is Ag-coated Cu with Kapton insulation (Accu-Glass Products model TYP32-15~\cite{accuglass}). Cabling of similar construction has been measured to be radiopure~\cite{xenon100-materials}.
This final step will be different for the planned screener: 
the G10-based PCBs used in the prototype will be replaced
with (more expensive) Cirlex,\footnote{Cirlex is an adhesiveless sheet material made from DuPont's Kapton polyimide and offered by Fralock~\cite{fralock}.} 
a material widely used in dark matter experiments because of its radiopurity~\cite{radiopurity}, 
low outgassing rate, and well-known mechanical properties (see~{\it e.g.}~\cite{alon}), and low-radioactivity solder~\cite{solderpaper} will be used.

Radon daughter contamination of the wires during stringing will be
prevented by undertaking construction in a radon-abated
cleanroom~\cite{lrt2013richard}.

\section{Prototype MWPC test setup}

In this section, 
the experimental configuration used to test the energy response and operational stability of the prototype MWPC is reviewed.
The completed frame was placed inside a stainless-steel pressure vessel that was connected to a rudimentary gas-handling system, providing a controlled space
for the introduction of a drift gas. Electrical connections were made via SHV vacuum feedthroughs attached to the bottom of the vessel, enabling application
of an electric field and signal readout.  

\subsection{Gas handling}\label{gas:hand}

Ultimately, the planned screener will be operated with a neon-methane mixture.  Neon is preferred because it has no 
naturally-occurring long-lived radiocontaminants, making it an ideal medium for low-background screening.  
Additionally, its stopping power is low enough that the trigger MWPC can be
thick enough (1\,cm) to make assembly straightforward.
For tests with the prototype, the more common (and less expensive) P-10 gas---a 
mixture of 90\%~Ar and 10\%~CH$_4$---was chosen for the detection medium.  The average energy loss of an incident particle to create a single electron-ion pair
in this mixture is $W = 26$\,eV~\cite{alk67}.  Therefore, 
38.5$\pm$2.6 electron-ion pairs are created 
per keV of deposited energy.\footnote{The Fano factor in P-10 gas is 0.17~\cite{alk67}.}

The gas-handling system consisted of a gas cylinder, an input manifold, and a turbo and roughing pump.  
The pressure vessel was first pumped down to a vacuum between $10^{-3}$--$10^{-5}$\,Torr before filling with room-temperature P-10 gas (purchased premixed).
The P-10 mixture was
allowed to flow freely into the chamber and was monitored with an analog relative-pressure gauge.  Flow was terminated (by hand)
once a slight overpressure of 0.5\,psig was achieved, thus minimizing contamination of the detection gas due to laboratory air leaking
into the pressure vessel.
Since the gas gain varies only linearly with pressure, this 
somewhat crude introduction of P-10 gas was sufficient to characterize the MWPC response; as shown in Fig.~\ref{gainvtime}, the gain varied by (at most) a few percent due to the 
resulting pressure variation.  Eventually, an MKS 902 piezoelectric transducer~\cite{mks} was added to monitor 
pressure to an accuracy of 1\%, permitting the measurements discussed in \S\ref{det:char}.

The gas-handling system for the planned screener will include measures
to avoid radon contamination~\cite{lrt2013ray}, reducing deposition
of radon daughters in the screener (especially on the MWPC wires), but
such measures were not taken for the tests reported here.

\subsection{Readout electronics}\label{readout:elec}
\begin{figure}[!tb]
\begin{center}
\includegraphics[width=5.95in]{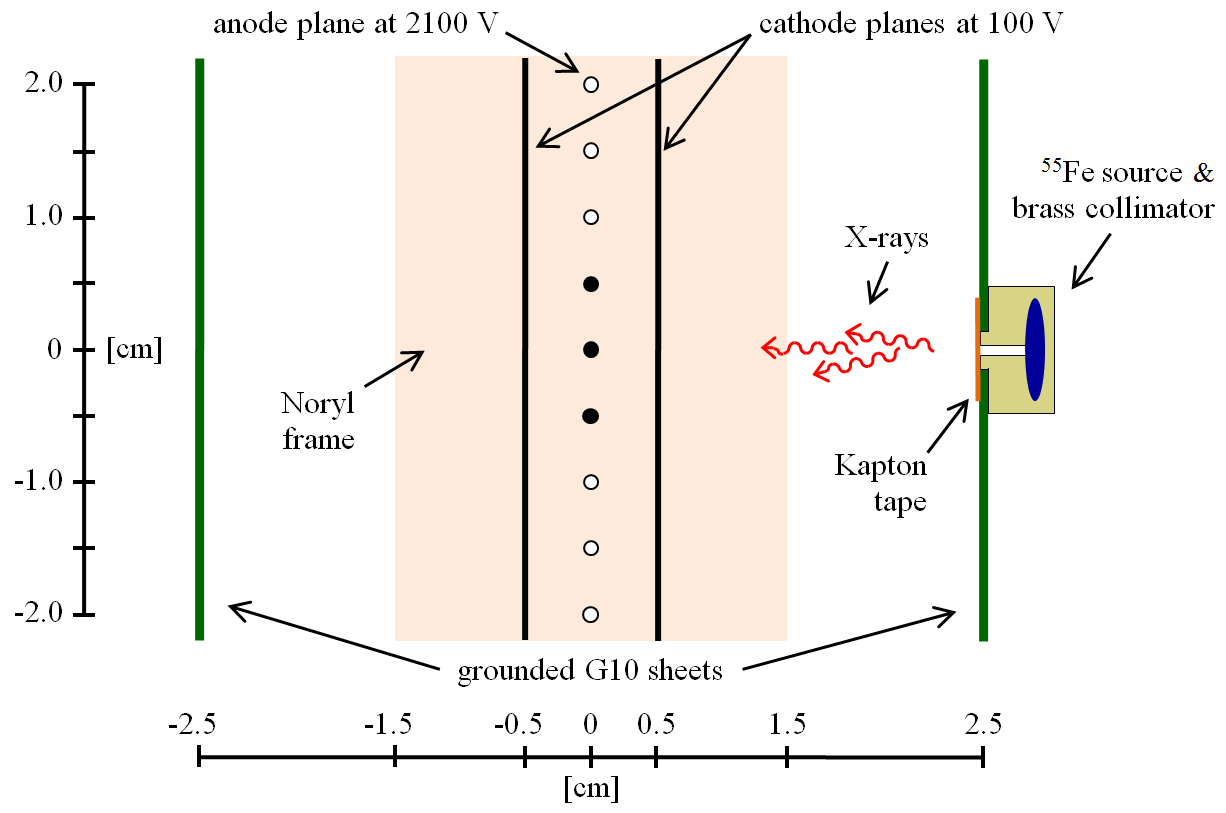}
\caption{Diagram of the drift-field and $^{55}$Fe-source configuration used to test the prototype MWPC,
highlighting the center of its crossed wire planes and drawn to scale.  The three central anode wires (dark dots) were read out
as a single channel, while the others (light dots) were biased but otherwise uninstrumented.  The shaded region corresponds to the 
Noryl frame's 3\,cm thickness.  Except for the Fig.~\protect\ref{spec}\ 
spectrum, all tests were made with the copper-clad G10 sheets attached to the 
Noryl frame via 1-cm thick nylon spacers, yielding a 2\,cm drift region for ionization electrons created between the Kapton tape and right-side
cathode wires (as shown).  The data featured in Fig.~\protect\ref{spec}\ 
were obtained with an earlier configuration in which the G10 sheets were bolted 
to the frame without spacers, resulting in a 1\,cm drift region.  This difference should have little to no effect on the energy response.}\label{schem}
\end{center}
\end{figure}

A simple single-channel data acquisition (DAQ) was used to test the performance of the MWPC.  The 
three central anode wires were ganged together (on the PCBs described in \S\ref{wire:string}) and read out as a single channel.  The remaining 
anode wires were held at the same potential---ranging from 1900--2250\,V for most tests---but were otherwise uninstrumented. During 
standard operations, the anode was biased to 2100\,V, while the cathode planes were kept at a small,  
100\,V potential to create a 2\,cm drift region between the cathode and the grounded G10 sheets bolted to the MWPC frame
(as indicated in Fig.~\ref{schem}).

The anode and cathode potentials were maintained by a Bertan Model 375P high-voltage power supply~\cite{bertan}, 
designed specifically for MWPC operation.  Since the gas gain is proportional to 
$(\lambda^\lambda)$, 
the voltage source must be extremely stable to prevent significant gain variations.
The high voltage was additionally conditioned by a two-stage low-pass 
filter with a 72\,Hz cut-off frequency.  All MWPC wires were biased with one end floating.  

A Cremat CR-111~\cite{cremat} two-stage charge amplifier was used to shape and amplify signals from the central three anode wires.  
Its first stage is a high-gain integrating amplifier with feedback resistance $R_{\mathrm{f}} = 10$\,M$\Omega$ and feedback capacitance $C_{\mathrm{f}}=15$\,pF,
yielding a 150\,$\mu$s output-pulse decay time.
The second stage is a low-gain amplifier intended to drive an output 
signal through a coaxial transmission line.
The rise time at the CR-111 output (3\,ns according to the manufacturer) is found to be faster than the digitization rate even with the additional input capacitance of the MWPC and so has no impact on the analyses. Similarly, the CR-111 gain (0.13\,V/pC according to the manufacturer) has not been verified experimentally, as its precise value is not critical; it is used only for confirmation of the Diethorn relation (\S4.2 and Fig.~\ref{diethornplot}).

A National Instruments PCI-5105 ADC~\cite{ni} was used to digitize the CR-111 output with 12-bit resolution at a rate of 250\,kS/s.  
The PCI-5105 was operated in continuous read digitization mode.\footnote{The PCI-5105 offers a triggered mode, but the software-limited live time was unacceptable.}
One-second-long 
traces were acquired starting at random times, and 
pulses were identified by searching for peaks in the time derivative of the traces.  Pulse heights were estimated from the peak values in the undifferentiated time stream of the 
identified pulses. 
To reduce high-frequency noise, a 6$^\mathrm{th}$-order 
Butterworth filter with a 10\,kHz cut-off frequency was applied to the raw traces prior to these steps.  This technique was used to measure pulse-height spectra, study pulse pile-up,
and to estimate absolute rates.  

Electronic noise was estimated from pulse-free sections of the same one-second-long, digitally filtered traces, yielding
$\sim$2\,mV FWHM.  This corresponds to $\sim$10$^{5}$ electrons FWHM at the CR-111 input, 
substantially poorer than the 1500--2500 electrons FWHM expected for the CR-111 from the manufacturer specifications
and allowing an input capacitance range of 0--120\,pF. 
As discussed in \S\ref{eresgain}, even this high noise is subdominant compared to Fano and avalanche
statistics and thus does not impede a demonstration that the MWPC provides near 
statistics-limited energy resolution, which was the goal of this effort.

\begin{figure}[!tb]
\begin{center}
\includegraphics[scale=1.0]{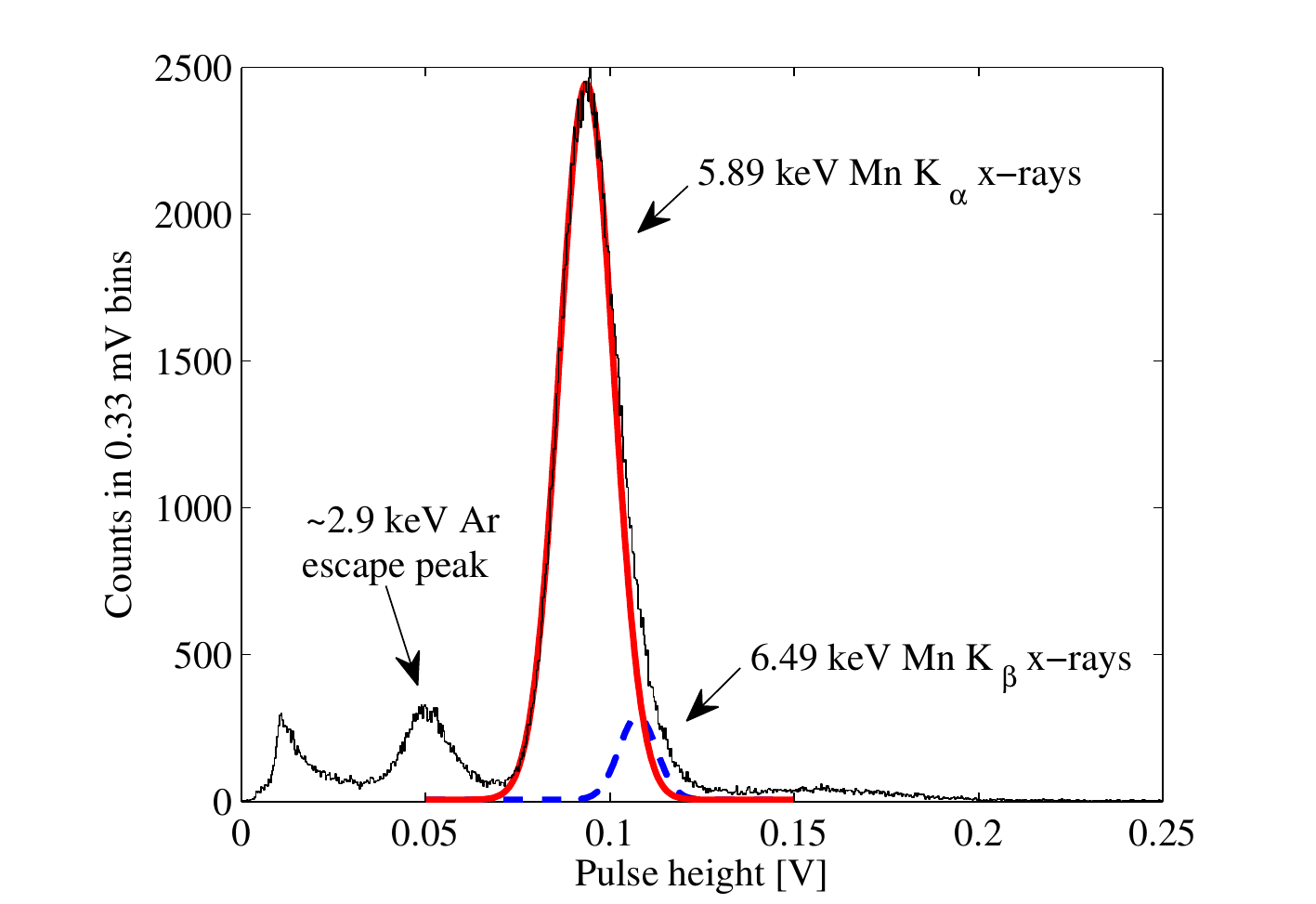}\llap{\makebox[7.25cm][l]{\raisebox{3cm}{\includegraphics[scale=0.35]{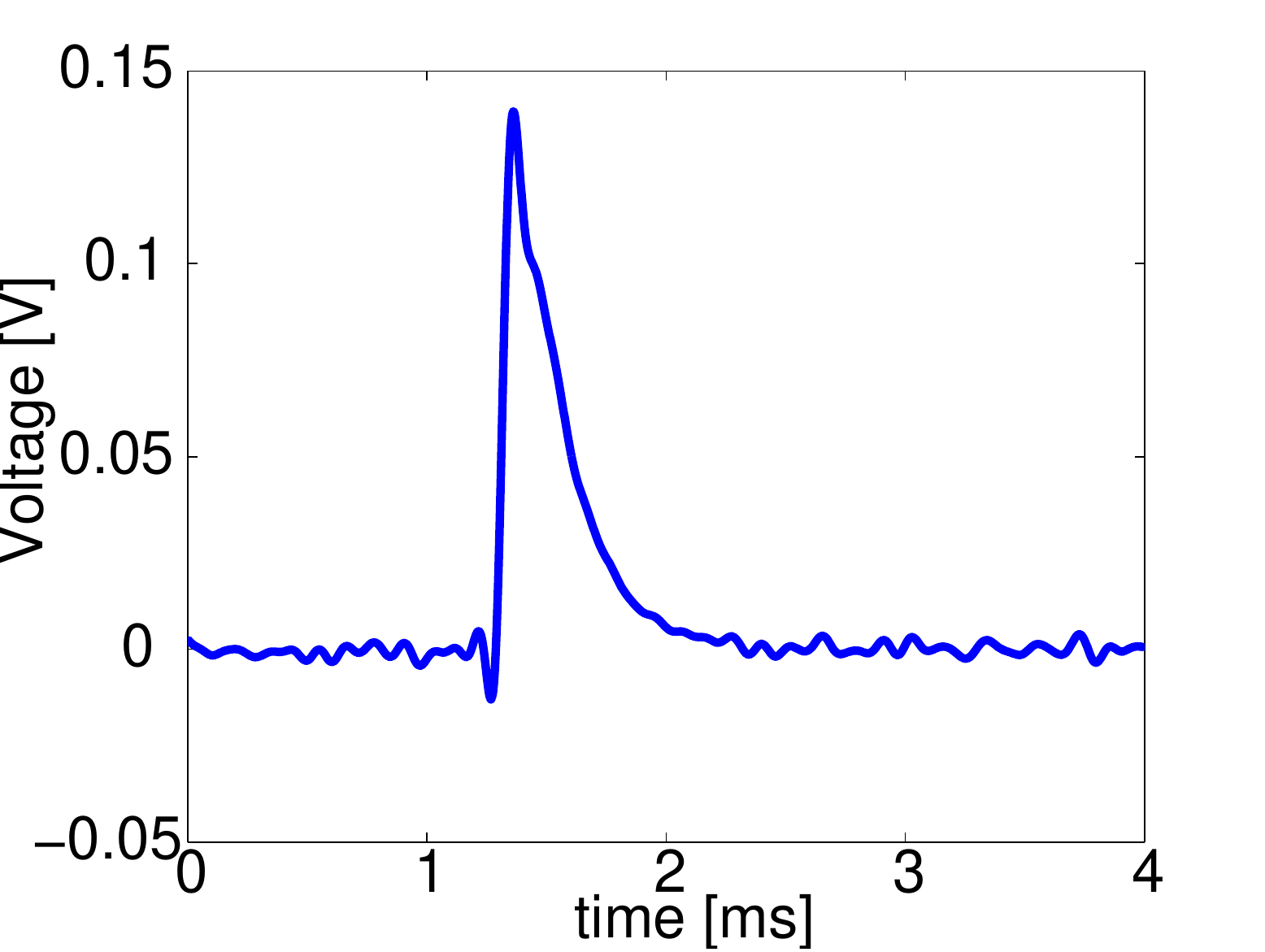}}}}
\caption{Pulse-height spectrum recorded with the prototype MWPC when exposed to an \fe\ source at  anode and cathode potentials set to 2100 and 100\,V, respectively.  The Mn K-shell x-rays are observed as a 
single peak with the addition of an $\sim$2.9\,keV Ar escape peak. The sources of the 0.01\,V peak and the broad hump 
in the 0.13--0.20\,V region are not known with certainty,
but they have no impact on the K-shell peak fits.  A combination of two Gaussians and a linear background was fit to
the region with pulse heights $\gtrsim$0.07\,V, where the means of the two Gaussians were constrained to have the expected 6.49-to-5.89 ratio.
This hypothesis has a $\chi^{2}$/d.o.f. = 502/487 and yields a FWHM resolution of 15.8\% for the 5.89\,keV peak, close to the minimum
possible value of 12--13\% ({\it cf.}~\S4.1 and appendix~A).  The best-fit Gaussians for the 5.89\,keV (solid) and 6.49\,keV (dashed) peaks are shown independently without the linear
background.  Inset: A typical $\sim$6\,keV pulse taken under the same conditions following application of a digital low-pass filter.}\label{spec}
\end{center}
\end{figure}

\section{MWPC performance}
To assess the performance of the prototype MWPC, several tests were conducted using x-rays from an \fe\ source.
Energy resolution and gain were measured as functions of voltage, pressure, and time, and estimates of the Diethorn parameters $\Delta V$ and $E_\mathrm{min}$ were extracted.

\subsection{Energy resolution and gain}\label{eresgain}
\fe\ decays via electron capture with a half-life of 2.73 years, producing Mn K-shell x-rays at 5.89 and 6.49\,keV.  The 5.89\,keV x-rays
have almost an order of magnitude larger branching fraction than the 6.49\,keV x-rays.  Further, in gaseous 
detectors the two energies tend to be indistinguishable due to the relatively large statistics-limited energy resolution.
The source activity at the time of the measurements was 26.5\,$\mu$Ci.  
Its active region is $\sim$3\,mm in diameter.  The choice 
of argon as the detection gas also introduces a 
2.9\,keV escape peak to the spectrum.  
When an x-ray ionizes a K-shell electron in Ar, the Ar atom emits an Ar K-shell x-ray with an 
energy of 3\,keV due to the filling of the K-shell vacancy. 
This 3\,keV x-ray has a longer absorption length than a higher-energy x-ray and so may escape the detection region, resulting in a peak 3\,keV below that of the primary x-ray. This is actually a collection of lines, making it difficult to fit for the energy resolution. 

The \fe\ source was centered on the MWPC immediately outside the drift region external to the 
copper-clad G10 ground plane ({\it cf.}\ Fig.~\ref{schem}).  The source was affixed to a 3.2-mm thick gold-plated copper collimator
with a 1-mm diameter circular aperture.  The front of the collimator was covered with a thin layer of 
Kapton tape to prevent electrons created outside the drift region from 
entering it. 
Note that the source and G10 sheet would not be installed during low-background operation of the planned screener.

\begin{figure}[!tb]
\begin{center}
\includegraphics[scale=0.55]{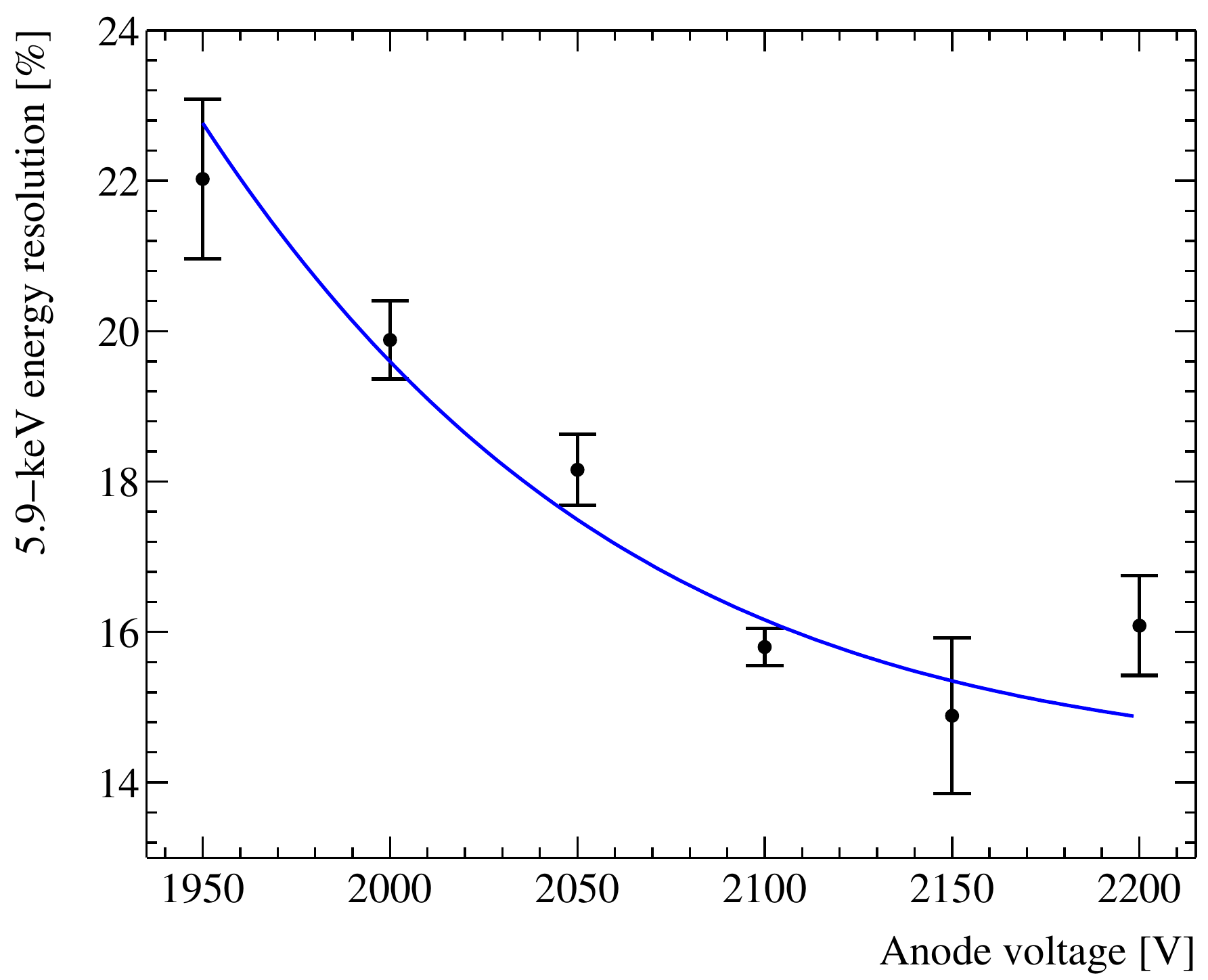} 
\caption{Measured FWHM energy resolution of the 5.89\,keV peak as a function of voltage.
The resolution degrades with decreasing bias voltage due to an increasing contribution
from a gain-dependent term in the relative resolution function ({\it cf.}\ Eq.\,4.1).  
The data (dots with 1$\sigma$ error bars) were taken in 1-minute intervals except at 2100\,V, for which a 20-minute acquisition was performed.  
The best-fit estimate ($\chi^{2}$/d.o.f. $= 8.5/4$) for the FWHM intrinsic resolution from Fano and avalanche statistics 
is $14.2\pm0.8$\%.}\label{fig.eresvsv}
\end{center}
\end{figure}

To search for the 5.89\,keV x-rays, a voltage scan was performed near an anode potential of 2100\,V.
These scans were taken both with and without the source present to 
confirm that any spectral features could indeed be attributed to the \fe\ source.
Once the peak was located, the voltage was tuned to bring the x-ray and escape peaks well above the noise threshold.  
Figure~\ref{spec} shows the pulse-height spectrum of an $^{55}$Fe-source run taken with anode and cathode potentials of 2100 and 100\,V, respectively.  
The Mn K-shell x-rays are observed along with an $\sim$2.9\,keV escape peak.  The best-fit mean of the 5.89\,keV peak was used to 
calibrate the energy scale and yields a gain of $\sim$10$^{4}$.

The 5.89\,keV peak fit featured in Fig.~\ref{spec} yields a 
FWHM energy resolution of 15.8\%.
Figure~\ref{fig.eresvsv} 
demonstrates that
the resolution improves 
with increasing gas gain (anode voltage) 
in the manner one would expect for the quadrature sum of a term due to Fano and avalanche statistics, $\mathrm{FWHM}_0$, and a term, $\mathrm{FWHM}_\mathrm{other}$, whose contribution
decreases with increasing gas gain (due to, {\it e.g.}, electronic noise):
\begin{equation}
\mathrm{FWHM}=\sqrt{\mathrm{FWHM}_0^2+\frac{\mathrm{FWHM}_\mathrm{other}^2}{\bar{G}^2}},
\end{equation}
where $\bar{G}$ is the pulse-height gain as a function of anode voltage normalized to the gain at 2100\,V. 
The values obtained at 5.89\,keV are $\mathrm{FWHM}_0 = 14.2\pm0.8$\% 
and $\mathrm{FWHM}_\mathrm{other} = 7.6\pm1.0$\%. 
As discussed in appendix~\ref{sec:appA}, the former is close to the minimum resolution of 12--13\% 
from Fano and avalanche statistics expected for our chamber configuration at 5.89\,keV.
The good agreement of the measurement with this value suggests that
any non-statistical contribution to $\mathrm{FWHM}_0$ is
smaller than the measurement precision.
The electronic-noise contribution is $\mathrm{FWHM}_\mathrm{elec} \approx 2.2\%$ based on the electronic
noise of $\sim$2\,mV FWHM from \S\ref{readout:elec} and the observed 5.89\,keV pulse height of 90\,mV, and 
thus $\mathrm{FWHM}_\mathrm{other}$ is appreciably larger than expected from electronic noise alone.
However, during typical operation for beta screening, the energy per 
5-mm cell 
will be approximately 1\,keV, at which energy 
the theoretical statistics-limited resolution will scale by $1/\sqrt{E}$ to approximately 30\% FWHM, likely rendering any such 
non-statistical contribution, as well as the electronic noise, 
negligible. 
Furthermore, 
any such non-statistical and electronic noises 
will 
decrease with upgrades to the electronics, gas-handling system, and a position-corrected analysis.

\begin{figure}[!tb]
\begin{center}
\includegraphics[scale=0.55,clip]{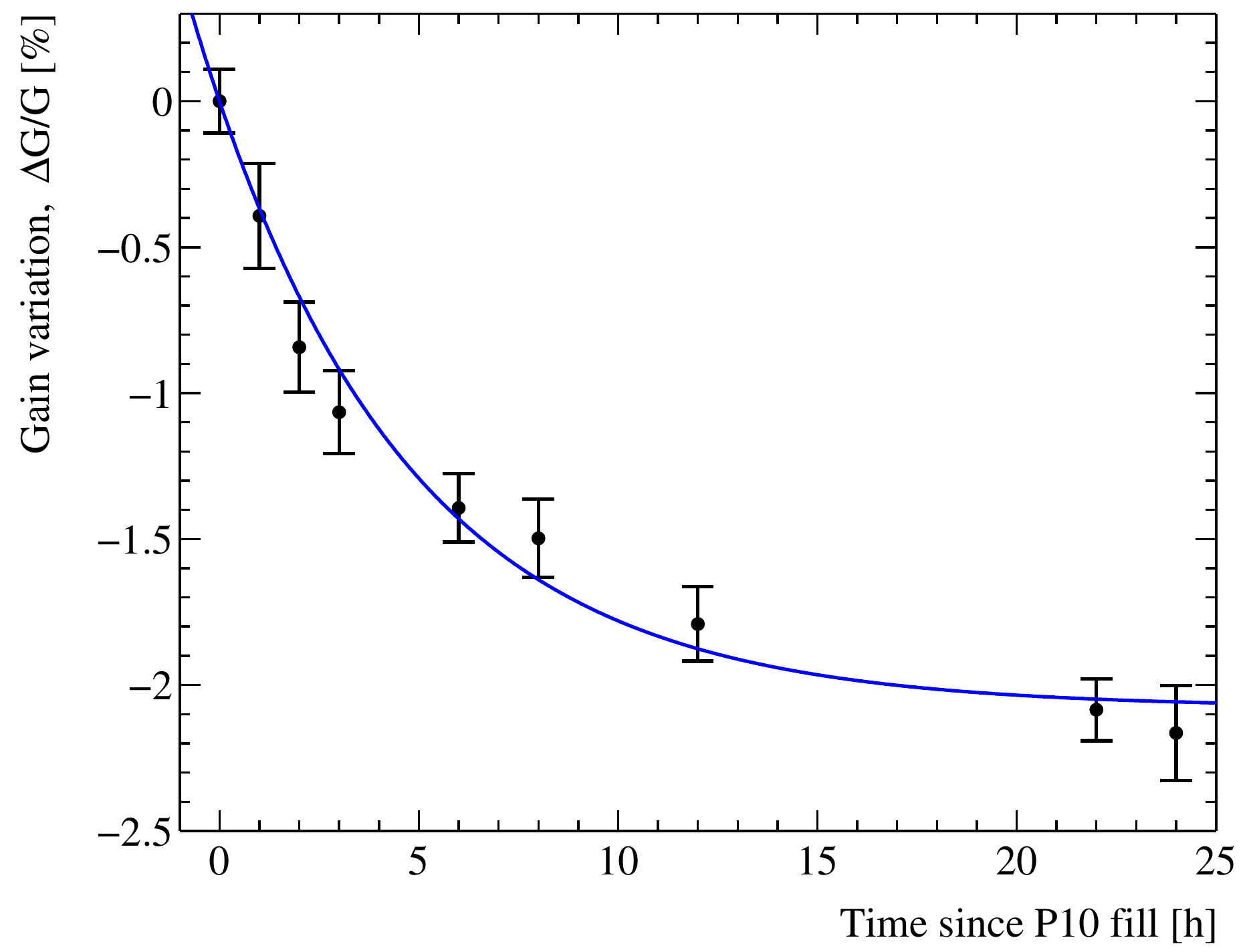} 
\caption{Stability of the gain ({\it vs.}\ time) following a P-10 gas fill. A series of $^{55}$Fe-source runs were taken over
the course of a day, with the means of the resulting 5.89\,keV peaks indicating a slight, few-percent degradation in gain (dots with 1$\sigma$ error bars).
Attributed to drift-gas contamination due to outgassing from detector materials, the degradation is well-fit by a two-parameter exponential form
({\it cf.}\ Eq.\,4.2) with a $\chi^{2}$/d.o.f. = 4.6/7, an amplitude $A_{\mathrm{stability}} = -2.08\pm0.09$\%, and a $1/e$ decay time 
$\tau_{\mathrm{stability}} = 5.1\pm0.7$\,h.}\label{gainvtime}
\end{center}
\end{figure}

\subsection{Detector characterization}\label{det:char}
To assess the prototype's operational stability, several short $^{55}$Fe-source runs were recorded over the course of a day subsequent to a P-10 gas fill.
Figure~\ref{gainvtime} shows the resulting stability of the gain, measured by the location of the 5.89\,keV peak, as a function of time.  
The observed gain variation is well-fit by an exponential trend:
\begin{equation}
\frac{\Delta G}{G} = A_{\mathrm{stability}}\left(1 - e^{-t/\tau_{\mathrm{stability}}}\right),
\end{equation} 
where $t$ is time since the P-10 gas fill. The best-fit amplitude $A_{\mathrm{stability}} = -2.08\pm0.09$\%, and the best-fit decay constant 
$\tau_\mathrm{stability}=5.1\pm0.7$\,h.  The primary reason for the gain drift is attributed to 
outgassing from the various detector components.  For the planned screener, the detection gas will be continuously circulated through
a SAES MC190-902F MicroTorr purifier~\cite{saes} and a (custom-design) cooled-carbon radon trap~\cite{lrt2013ray} to help remove outgassed contaminants
and thus further improve gain stability.

\begin{figure}[!tb]
\begin{center}
\includegraphics[width=5.9595in]{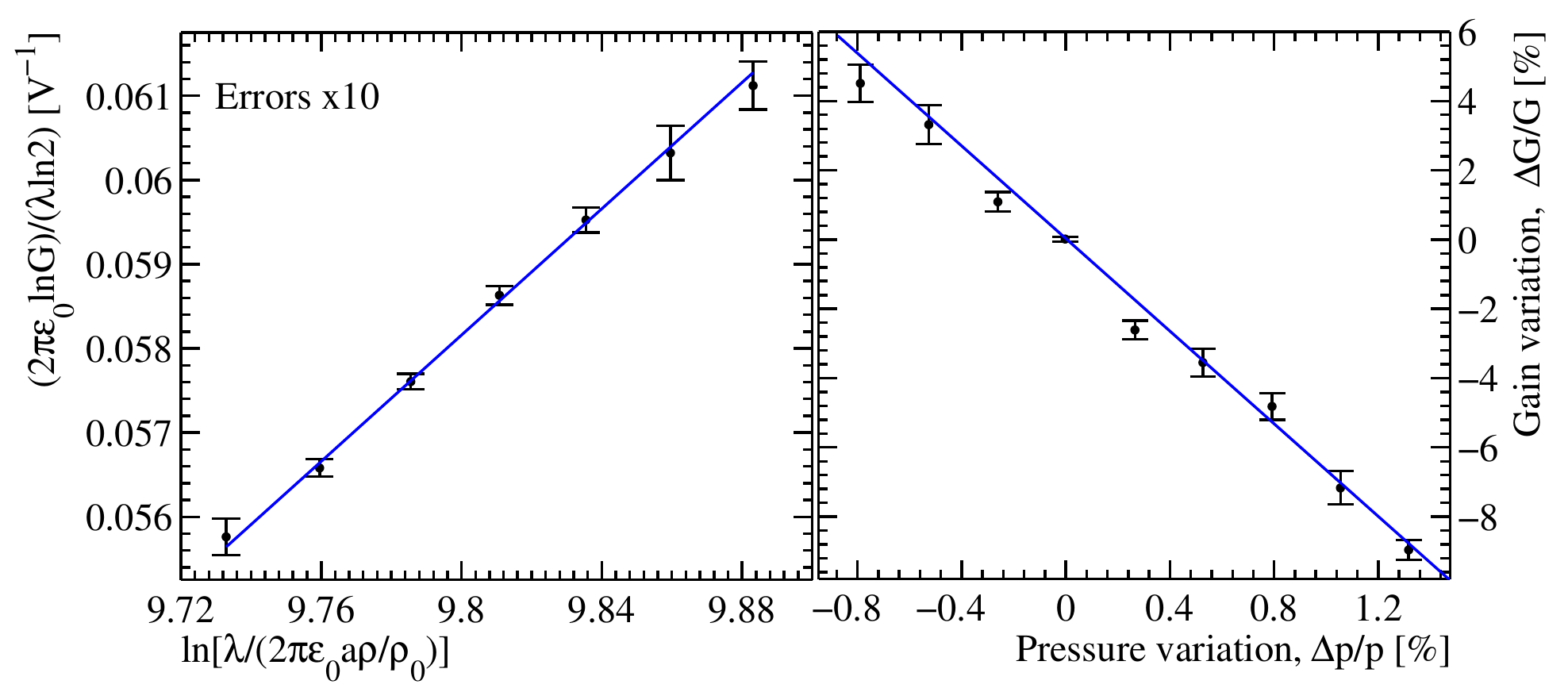} 
\begin{tabular}{c@{\hspace{3.2in}}c}
a)&b)
\end{tabular}
\caption{a) Test of the Diethorn formula for gas gain in a proportional chamber, where Eq.\,2.1 has been rearranged to yield a linear relationship with varying anode voltage (dots with exaggerated 1$\sigma$ error bars).
An estimate of the Diethorn parameters is extracted with a linear fit; $\Delta V = 26.7\pm0.6$\,V and $E_{\mathrm{min}} = 38.3\pm1.3$\,kV/cm, in agreement with expectation.
b) Gain variation, as measured with 5.89\,keV x-rays, as a function of pressure change (dots with 1$\sigma$ error bars).  The Diethorn formula predicts a slope of $-6.5\pm0.3$ that is
confirmed by a best-fit slope of $-6.7\pm0.3$.}\label{diethornplot}
\end{center}
\end{figure}

The Diethorn parameters $\Delta V$ and $E_\mathrm{min}$ were estimated with a fit to $(2\pi\epsilon_0\ln G)/(\lambda\ln2)$ 
as a function of $\ln|\lambda/(2\pi\epsilon_0a\rho/\rho_0)|$ ({\it cf.}\ Fig.~\ref{diethornplot}a), where the variation in the charge
density results from varying the high voltage.  This linear relationship allows for easy extraction of the Diethorn parameters; $\Delta V=26.7\pm0.6$\,V and 
$E_\mathrm{min}=38.3\pm1.3$\,kV/cm,\footnote{Assuming the nominal 0.13\,V/pC gain of the Cremat CR-111 amplifier~\cite{cremat}.} with expectations of $23.6\pm5.4$\,V and $48\pm3$\,kV/cm, respectively~\cite{hen}.\footnote{Historically, there is a wide range of measured $E_\mathrm{min}$ values that encompass 
this estimate~\cite{pddc}.}  All uncertainties are statistical at 1$\sigma$ confidence.

An additional check on the performance of the MWPC is to vary the pressure and monitor the corresponding change in gain.  
Figure~\ref{diethornplot}b shows this measurement, again using the location of the 5.89\,keV peak to track the gain.  
This linear relationship between gain and pressure is expected to have a slope of $-6.5\pm0.3$~\cite{pddc}.  The best-fit slope to the 
data in Fig.~\ref{diethornplot}b is $-6.7\pm0.3$.

Table~\ref{fitpar} compares the expected values of several characteristic MWPC parameters to those measured with 
the prototype.
Where appropriate, all parameters match to within statistical uncertainties, with $E_\mathrm{min}$ to within the spread of previous measurements~\cite{pddc}.  
The spread is most likely due to slight impurities in the detection gas.
As dependence of the gain on this parameter is somewhat muted due to a logarithm, the modest inconsistency with previous values is not a concern.

\section{Conclusion}
A prototype low-background MWPC for the detection of alphas and low-energy betas has been constructed and characterized.  This device performs 
well down to a few keV in deposited energy as shown by the measurement of x-rays produced from \fe\ decays.  The MWPC was found to respond 
with an energy resolution of 15.8\% at 5.89\,keV and a gas gain of $\sim$10$^{4}$, close to the expected statistical limit.  Further improvements to this resolution 
will come from planned electronics upgrades to include position information.  The gain is stable to $\sim$2\% for $>$1-day periods and is expected to 
have negligible instability with a planned upgrade to the gas-handling system that includes a continuous circulation loop with gas purification.
Further, gain variations due to pressure changes fall in line with expectation and can be corrected with accurate pressure monitoring.  
This prototype MWPC has met or exceeded all of its design goals.

\begin{table}[tb!]
\begin{center}
\caption{Comparison of several expected gas parameters with those measured with the prototype MWPC. 
As outgassing is hard to predict, there is no expectation for the two stability parameters. All uncertainties
are quoted at 1$\sigma$ confidence.}\label{fitpar}
\begin{tabular}{l|c|r@{$\pm$}l|r@{$\pm$}l|l}
\multicolumn{1}{c}{parameter} & \multicolumn{1}{c}{description} & \multicolumn{2}{c}{expected} & \multicolumn{2}{c}{measured} &\multicolumn{1}{c}{units}\\ \hline\hline
$\mathrm{FWHM}_0$                  &	intrinsic resolution @ 5.89\,keV	&\multicolumn{2}{c|}{12--13}& 14.2&0.8 & \% \\
$A_\mathrm{stability}$      &	gain-stability amplitude	&\multicolumn{2}{c|}{}& -2.08&0.09& \% \\ 
$\tau_\mathrm{stability}$   &	gain-stability $1/e$ decay time	&\multicolumn{2}{c|}{}&  5.1&0.7 & h \\
$E_\mathrm{min}$            &	Diethorn minimum electric field &   48&3   & 38.3&1.3 & kV/cm \\
$\Delta V$                  &	Diethorn electron-ion potential & 23.6&5.4 & 26.7&0.6 & V     \\
$\frac{\partial\left(\Delta G/G\right)}{\partial\left(\Delta p/p\right)}$ &	gain {\it vs.}\ pressure slope	& -6.5&0.3 & -6.7&0.3 &
\end{tabular}
\end{center}
\end{table}

\acknowledgments
This work was supported in part by the National Science Foundation (Grants No.\ PHY-0834453 and PHY-0855525), the 
Department of Energy HEP division, and  the University of Alberta Department of Physics and Faculty of Science.
The authors gratefully acknowledge the technical contributions of L.~Buda, J.~Hanson, 
G.\ Lachat, and P.\ Zimmerman, and the quality-control testing of J.~Roberts and M.~Mintskovsky.

\appendix 
\numberwithin{equation}{section}

\section{Intrinsic Energy Resolution} 
\label{sec:appA} 

Intrinsic proportional-chamber resolution at energy $E$ can be estimated using 
Eq.~6.24 in~\cite{knoll}:
\begin{equation}
\mathrm{FWHM}_{0} = 2.35\sqrt{\frac{W(F+b)}{E}},
\end{equation} 
where $W=26$\,eV is the average energy required
to form one electron-ion pair in P-10~\cite{alk67}, $F=0.17$ is the Fano factor for P-10~\cite{alk67}, and $b$ is the avalanche factor and depends not only on
the gas type but also on the chamber's pressure and anode-wire radius. We estimate $b\approx0.44$ following Alkhazov's model~\cite{alk70}, as summarized in~\cite{knoll,sipila,alk69}.
Alkhazov's model appears to agree well with experimental data for argon-methane chambers with gas gains between $\sim$100 and $10^{5}$ (see, {\it e.g.},~\cite{CandC}).
We start with $b=0.5$, calculated using Alkhazov's model for a gas gain of 100 and a pressure$\times$anode-radius product of 0.5\,Torr-cm~\cite{sipila}.  Using the figures
in~\cite{alk69} (also from Alkhazov's model), we scale down by 18\% to account for our chamber's smaller pressure$\times$anode-radius product ($\sim$0.01\,Torr-cm) 
and then up by 7.5\% to account for the larger gas gain ($\sim$10$^{4}$), yielding $b\approx0.44$ and a minimum $\mathrm{FWHM}_0$ 
at 5.89\,keV of $\sim$12\%.

Of course, Alkhazov's model has not
been fully verified as it is impossible to construct a perfect chamber.  The measurement in~\cite{sipila} of 13.2\% at 5.89\,keV is the lowest resolution using 
P-10 in the literature and is very close to the model's predicted 12.8\%.  Treating this measurement as an upper limit on the resolution due to Fano and avalanche statistics implies $b\leq0.55$ 
for the device used in~\cite{sipila}, which, if scaled with pressure$\times$anode-radius product and gas gain to our chamber's parameters as was done above, suggests
$b\lesssim0.5$.  To encompass this experimental uncertainty, a minimum expected $\mathrm{FWHM}_0$
of 12--13\% is used in this paper, corresponding to $b=0.44$--0.5.

\bibliographystyle{JHEP}
\bibliography{betacage_jinst}

\end{document}